\documentclass{ws-ijmpc}

\begin{document}

\markboth{Dami\'an H. Zanette} {Demographic growth and the
distribution of language sizes}

\catchline{}{}{}{}{}

\title{DEMOGRAPHIC GROWTH AND THE DISTRIBUTION OF LANGUAGE SIZES}

\author{DAMI\'AN H. ZANETTE}

\address{Consejo Nacional de Investigaciones Cient\'{\i}ficas y
T\'ecnicas\\
Centro At\'omico Bariloche and Instituto Balseiro\\
8400 San Carlos de Bariloche, R\'{\i}o Negro, Argentina.\\
zanette@cab.cnea.gov.ar}

\maketitle

\begin{history}
\received{Day Month Year}
\revised{Day Month Year}
\end{history}

\begin{abstract}
It is argued that the present log-normal distribution of language
sizes is, to a large extent, a consequence of demographic dynamics
within the population of speakers of each language. A two-parameter
stochastic multiplicative process is proposed as a model for the
population dynamics of individual languages, and applied over a
period spanning the last ten centuries. The model disregards
language birth and death. A straightforward fitting of the two
parameters, which statistically characterize the population growth
rate,  predicts a distribution of language sizes in excellent
agreement with empirical data. Numerical simulations, and the study
of the size distribution within language families, validate the
assumptions at the basis of the model. 
\end{abstract}

\ccode{PACS Nos.: 87.23.Ge, 89.65.Cd, 02.50.Ey}

\section{Introduction}

Statistical aspects of the evolution of languages have attracted, in
the last few years, a great deal of attention among physicists and
mathematicians.\cite{nowak,dhz,cancho,strog,schulze,schw,viviane,wich,x,oli0,t1}
One of the better established quantitative empirical facts about
extant languages is their size distribution, namely, the frequency
of languages with a given number of speakers. Naturally, explaining
the origin of this distribution is one the aimed goals of
mathematical modeling in this field.

Recent work has built up on variations of two basic models of
language evolution, Schulze's model\cite{schulze} and Viviane's
model,\cite{viviane} both of them mostly focused on the effects of
mutation of linguistic features which give rise to new languages,
and including the possibility of language extinction. These models,
however, disregard the fact that, over periods which are short as
compared with the typical time scales of language evolution, the
speakers of a given language can substantially vary in number just
by the effect of population dynamics. For instance, in the last five
centuries --a period which includes the culturally devastating
European invasion of the rest of the globe-- perhaps 50~\% of the
world's languages went extinct (among them, two thirds of the 2,000
preexisting native American languages\cite{camp}) or changed
drastically. In the same period, however, the world's population
grew by a factor of twelve or more.\cite{y1000}

Demographic effects have been very recently incorporated to a model
of language evolution by Tuncay,\cite{t1,t2} in the form of a
stochastic multiplicative model for population growth (see also Ref.
\refcite{sol}). With suitable tuning of its several parameters,
Tuncay's model is able to reasonably reproduce the observed
distribution of language sizes, as a result of numerical
simulations.\cite{t1} The complexity of the model --which, in its
full form,\cite{t2} includes population growth along with language
inheritance, branching, assimilation, and extinction-- makes it
however difficult to identify the specifc mechanism that shapes the
distribution of sizes.

In this paper, I  show that the empirical distribution of language
sizes can be accurately explained taking into account just the
effect of demographic processes. This possibility was already
pointed out, for the specific case of the languages of New Guinea,
by Novotny and Drozd.\cite{novotny} I propose a two-parameter
stochastic model where the populations speaking different languages
evolve independently of each other. During the evolution --which, in
the realization of the present model, is assumed to span 1,000
years-- language creation, assimilation, and extinction are
disregarded. This assumption does not discard mutations inside a
given language, which may lead to its internal evolution, but each
language preserves its identity as a cultural and demographic unit
along the whole period. The model is analytically tractable, and its
two parameters can be fitted {\it a priori} from empirical data.
Numerical simulations confirm the prediction that the distribution
is essentially independent of details in the initial condition --the
distribution of sizes ten centuries ago-- so that, in a sense, the
present distribution is the unavoidable consequence of just
demographic growth. The model is further validated by showing that
the distribution of sizes of languages belonging to a given family
has the same shape as the overall distribution. These results
strongly suggest that population dynamics is a necessary ingredient
in models of linguistic evolution.

\section{Present distribution of language sizes}

It is a well established empirical fact that the frequency of
languages with a given number of speakers $p$ is satisfactorily
approximated by a log-normal distribution.\cite{sutherland,schulze}
Accordingly, the distribution of language sizes as a function of the
log-size, $q=\ln p$,  is approximately given by a Gaussian,
\begin{equation} \label{Gauss}
Q(q) = (2\pi \theta^2)^{-1/2} \exp \left[ -(q-\bar q)^2 /2\theta^2
\right],
\end{equation}
where $\bar q$ and $\theta$ are, respectively, the mean value and
the mean square dispersion of $q$. The quantity $Q(q) dq$ gives the
fraction of languages with log-sizes in $(q,q+dq)$. Significant
departures from the log-normal distribution are limited to small
language sizes, up to the order of a few tenths
speakers.\cite{sutherland}

\begin{figure}[h]
\centerline{\psfig{file=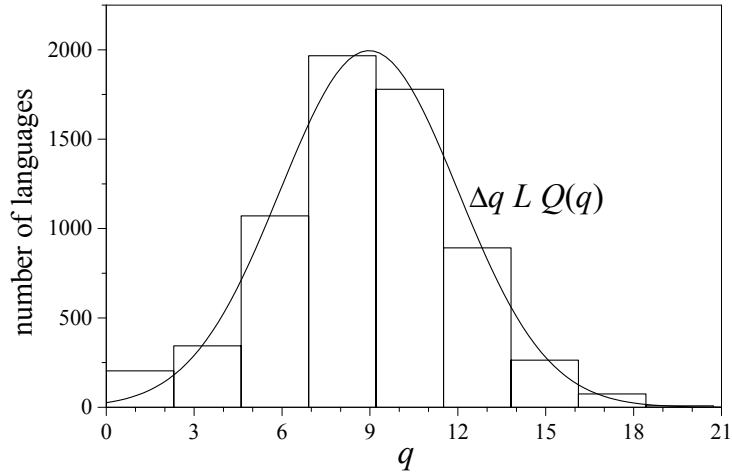,width=10cm}}  \vspace*{8pt}
\caption{The histogram shows the number of languages whose sizes $p$
have logarithms $q=\ln p$ in the corresponding bins, according to
the Ethnologue statistical summaries.{\protect \cite{ethnologue}}
The curve is the Gaussian distribution $Q(q)$ of Eq. (\ref{Gauss})
with the parameters of Eq. (\ref{fit}). For comparison with the
histogram, $Q(q)$ is multiplied by the total number of languages,
$L=6,604$, and by the bin width $\Delta q = \ln 10$. Note the
deviation from normality in the leftmost column of the
histogram.\protect \cite{sutherland} \label{f1}}
\end{figure}

Ethnologue statistical summaries,\cite{ethnologue} whose data
correspond to collections done mostly during the 1990s, list the
number of languages with sizes within decade bins (1 to 9 speakers,
10 to 99 speakers, 100 to 999 speakers, and so on), and give the
number of speakers within each bin. The total number of languages in
the list is $L = 6,604$, accounting for an overall population a
little above $5.7 \times 10^9$ speakers. The sizes of $308$
languages of the database are unknown. In terms of the distribution
$Q(q)$, the number of languages in the bin between $10^k$ and
$10^{k+1}$ speakers is
\begin{equation}
L_k =L \int_{k \ln 10}^{(k+1)\ln 10} Q(q) \ dq,
\end{equation}
while the total population speaking the languages in the same bin is
\begin{equation}
p_k =L \int_{k \ln 10}^{(k+1)\ln 10} \exp(q) \ Q(q) \ dq.
\end{equation}
The values of $L_k$ and $p_k$ provided  by the Ethnologue
statistical summaries can thus be used to estimate the parameters
$\bar q$ and $\theta$ in the distribution $Q(q)$ of Eq.
(\ref{Gauss}). Least-square fitting yields
\begin{equation} \label{fit}
\bar q = 8.97 \pm 0.15 , \ \ \ \ \ \theta = 3.04 \pm 0.09 .
\end{equation}

Figure \ref{f1} shows, as a histogram over the variable $q$,
Ethnologue's data for $L_k$. The curve is a plot of the function
$LQ(q)$ with the parameters of Eq. (\ref{fit}), whose integral over
the histogram bins approximates the empirical values of $L_k$. For
easier comparison with the histogram, $LQ(q)$ is further multiplied
by the bin width $\Delta q = \ln 10$. The aim in the following is to
provide a model which explains the fitted distribution $Q(q)$.

\section{Demographic evolution of language sizes} \label{sect3}

The log-normal shape of the distribution of language sizes suggests
that a multiplicative stochastic process is at work in the evolution
of the number of speakers of each language. This is in turn
consistent with the hypothesis that, over sufficiently long time
scales, the population speaking a given language evolves
autonomously, driven just by demographic processes.\cite{novotny}

Let $p_t^{(j)}$ be the number of speakers of language $j$ at time
$t$, and assume that at time $t+1$ --one year later, say-- the
population has changed to\cite{t1}
\begin{equation} \label{pt0}
p_{t+1}^{(j)} = \alpha_t^{(j)} p_t^{(j)},
\end{equation}
where the growth rate $\alpha_t^{(j)}$ is a positive stochastic
variable drawn from some specified distribution. As shown below, the
mean value and the mean square dispersion over this distribution can
be estimated from empirical data. In terms of the initial population
$p_0^{(j)}$, the number of speakers at time $t$ is
\begin{equation} \label{pt}
p_t^{(j)} = p_0^{(j)} \prod_{s=0}^{t-1} \alpha_s^{(j)} .
\end{equation}
I suppose now that, during the whole $t$-step process, the
distribution of the growth rate $\alpha_t^{(j)}$ is (i) the same for
all languages, and (ii) does not vary with time. Moreover, (iii) no
language is created or becomes extinct. Admittedly, these are rather
bold assumptions for the world's history during the last 1,000
years. However, in view of the lack of reliable data over such
period, they are at least justified by the sake of simplicity.

I identify the evolution of the world's languages as $L=6,604$
realizations of the multiplicative stochastic process (\ref{pt0}).
By virtue of  assumption (i), all the realizations are statistically
equivalent. In this interpretation, the present distribution of
log-sizes, given by Eqs. (\ref{Gauss}) and (\ref{fit}), is the
probability distribution for the variables $p_t^{(i)}$ obtained from
those realizations. Thus, my aim is to quantitatively relate the
distribution $Q(q)$ to the outcome of the stochastic process.

The total population $P_t$ at time $t$ is
\begin{equation}
P_t = \sum_{j=1}^L p_t^{(j)} = \sum_{j=1}^L p_0^{(j)}
\prod_{s=0}^{t-1} \alpha_s^{(j)} .
\end{equation}
Averaging this expression over realizations of the stochastic
variable $\alpha_t^{(j)}$, and assuming that the growth rate is not
self-correlated in time, we find $\langle P_t \rangle = \langle
\alpha \rangle^t P_0$, where $\langle \alpha \rangle$ is the mean
growth rate and $P_0$ is the initial total population. In order to
apply this analysis to the world's population in the last ten
centuries, let us take $P_0=3.1 \times 10^8$, which is the estimated
population by the year 1000.\cite{y1000} Ascribing the total
population accounted for by Ethnologue's data to the year 2000, and
associating this number with the population averaged over
realizations of the growth rate, we have $\langle P_t \rangle = 5.7
\times 10^9$ and $t=10^3$. This makes it possible to evaluate the
mean growth rate per year as
\begin{equation} \label{alfaav}
\langle \alpha \rangle = \left(\langle P_t \rangle /P_0
\right)^{1/t}  \approx 1.0029.
\end{equation}

To evaluate the dispersion of the growth rate, it is useful to
introduce  its relative deviation with respect to the average,
$\delta_t^{(j)}$, as
\begin{equation}
\alpha_t^{(j)}= \langle \alpha \rangle \left[ 1+ \delta_t^{(j)}
\right] .
\end{equation}
The average value and the mean square dispersion of the deviation
$\delta_t^{(j)}$ are, respectively,
\begin{equation}
\langle \delta_t^{(j)} \rangle = 0 , \ \ \ \ \ \sigma_\delta \equiv
\langle  \delta_t^{(j)2} \rangle^{1/2} = \sigma_\alpha/ \langle
\alpha \rangle,
\end{equation}
where $\sigma_\alpha$ is the mean square dispersion of the growth
rate. Assuming that $\delta_t^{(j)}$ is always sufficiently small to
approximate $\ln ( 1+ \delta_t^{(j)}) \approx \delta_t^{(j)}-
\delta_t^{(j)2}/2 $, Eq. (\ref{pt}) can be rewritten for the
log-size $q_t^{(j)}= \ln p_t^{(j)}$ as
\begin{equation} \label{qt}
q_t^{(j)} = q_0^{(j)}+t \ln \langle \alpha \rangle +
\sum_{s=0}^{t-1} \left[ \delta_s^{(j)}-\delta_s^{(j)2}/2 \right] .
\end{equation}
This equation shows explicitly that, besides the deterministic
growth given by the term $t \ln \langle \alpha \rangle $, the
evolution of the logarithm of the population speaking a given
language is driven by an additive stochastic process. Thus, by
virtue of the central limit theorem,\cite{CLT} the distribution
$Q(q)$ must converge to a Gaussian like in Eq. (\ref{Gauss}),
starting from any distribution of initial log-sizes $q_0^{(j)}$. For
the time being, however, the question remains whether the times
relevant to the process are enough to allow for the development of
the Gaussian shape and, in particular, to suppress any effect
ascribable to a specific initial distribution.

Unfortunately, the initial sizes $p_0^{(j)}$ --the number of
speakers of each language 1,000 year ago-- are not known for most
languages. However, their effect on the present size distribution
can be readily assessed. Averaging Eq. (\ref{qt}) over realizations
of the stochastic process and over the distribution of initial
log-sizes --and always assuming that the deviations $\delta_t^{(j)}$
are small--  yields, for the mean square dispersion of $q_t^{(j)}$,
\begin{equation} \label{sq2}
\sigma_q^2 = \sigma^2_0+ \sigma_\delta^2 t,
\end{equation}
where $\sigma_0$ is the mean square dispersion of the initial
log-sizes. The empirical estimation for $\sigma_q$ is the value of
$\theta$ given in Eq. (\ref{fit}). In turn, an upper bound can be
given for $\sigma_0$, as the maximal mean square dispersion in the
log-size distribution of $L=6,604$ languages with a total population
of $P_0=3.1 \times 10^8$ speakers, and at least one speaker per
language. This maximal dispersion is obtained with $L-1$ languages
with exactly one speaker, and just one language with the remaining
$P_0-L+1$ speakers. Clearly, this is an unlikely distribution for
the languages 1,000 years ago, but represents the worst-case
instance, with the largest contribution of the initial condition to
the present dispersion of log-sizes. In this extreme situation, the
estimation for the initial mean square dispersion is $\sigma_0^2
\approx L^{-1 } \ln^2 P_0 \approx 0.058$. Meanwhile, according to
Eq. (\ref{fit}), $\sigma_q^2 \approx 9.2$. In the right-hand side of
Eq. (\ref{sq2}), therefore, the largest contribution by far is given
by the second term, and that of the initial distribution is
essentially negligible. This makes it possible to calculate the mean
square dispersion of the deviations $\delta_t^{(j)}$ as
\begin{equation} \label{sigmad}
 \sigma_\delta \approx t^{-1/2} \theta \approx 0.096
\end{equation}
for $t=10^3$, thus completing the statistical characterization of
the growth rate  per year, $\alpha_t^{(j)}$. Note that this value of
$\sigma_\delta$ is in agreement with the assumption of small
relative deviations in the growth rate.

Summarizing the results of this section, I have argued that the
present log-normal distribution of language sizes can be seen as the
natural consequence of population dynamics driven by a stochastic
multiplicative process, Eq. (\ref{pt0}), where the evolution of each
language is interpreted as a realization of the process. Using data
on the total population growth during the last 1,000 years --a
period over which I neglect language birth and death-- and fitting
only one parameter ($\sigma_\delta$) from the distribution itself, I
was able to statistically characterize the growth rate per year
which explains the present distribution, giving its mean value and
mean square dispersion. Also, I have advanced that the dispersion of
language sizes ten centuries ago has essentially no effect on its
present value. It is now useful to validate these conclusions with
numerical realizations of the model, and with applications within
language families.

\section{Validation of the model}

\subsection{Numerical results}

In this section, I present results for series of $L=6,604$ numerical
realizations of the multiplicative stochastic process (\ref{pt0}).
The mean value and the mean square dispersion of the growth rate
$\alpha_t^{(i)}$ are fixed according to the values estimated in
Section \ref{sect3}, Eqs. (\ref{alfaav}) and (\ref{sigmad}). In
order to speed up the computation, individual values of
$\alpha_t^{(i)}$ are drawn from a square distribution centered at
$\langle \alpha \rangle$, with a width which insures the correct
mean square dispersion. In agreement with my main assumptions, I
avoid the possibility that languages die out by replacing the
absorbing boundary at $p=1$, below which a language should become
extinct, by a reflecting boundary.

\begin{figure}[h]
\centerline{\psfig{file=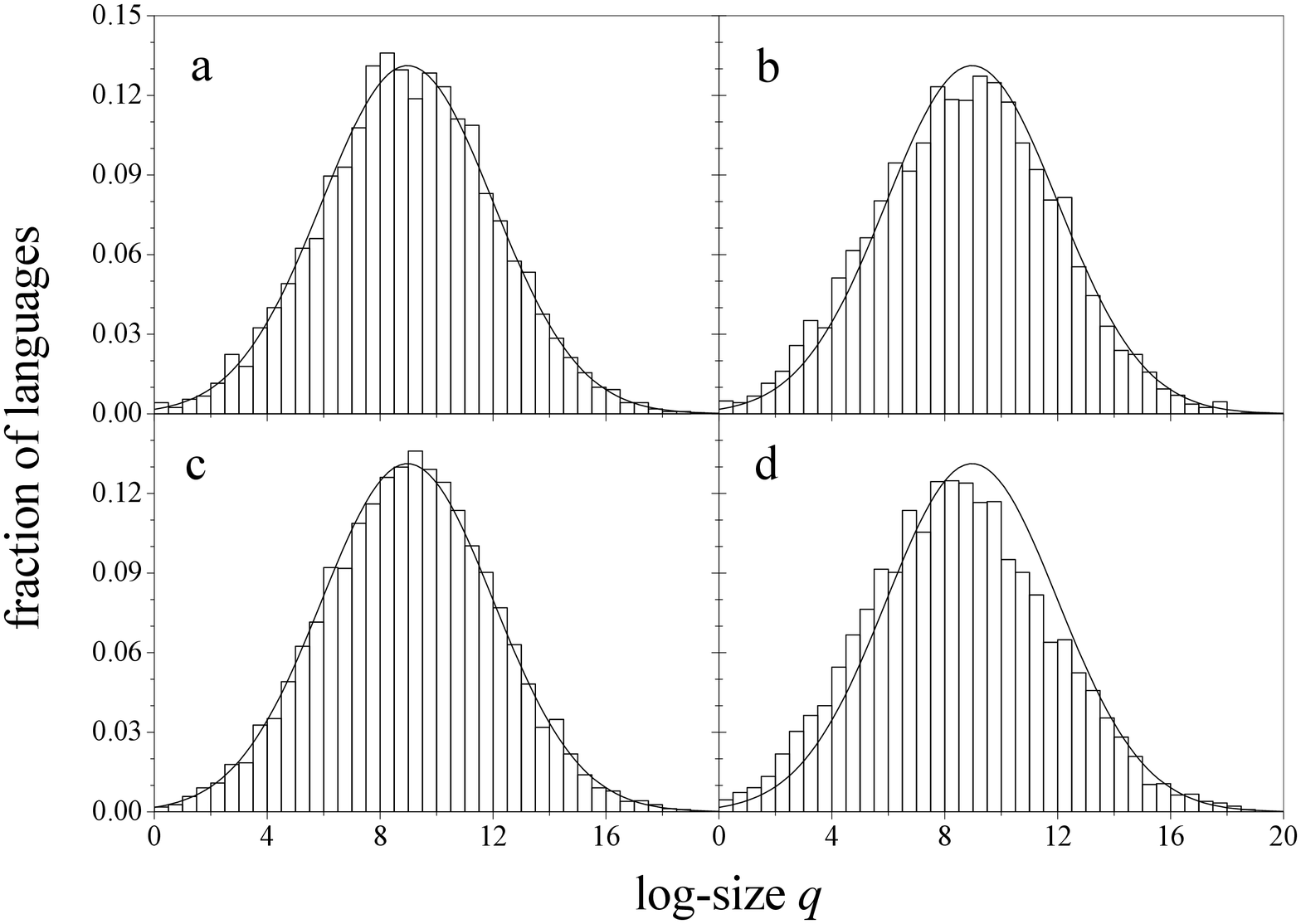,width=10cm}} \vspace*{8pt}
\caption{Normalized histograms for $L=6,604$ language sizes after
$10^3$ steps of the stochastic process (\ref{pt0}), starting from
the four initial conditions described in the text. Curves stand for
the expected Gaussian distribution, Eq.  (\ref{Gauss}), with the
parameters of Eq. (\ref{fit}).\label{f2}}
\end{figure}

In view of the discussion in the previous section, the convergence
of the distribution of log-sizes to a Gaussian is guaranteed by the
central limit theorem. The emphasis in the simulations is thus put
on the possible effects of the distribution of initial sizes
$p_0^{(i)}$. Figure \ref{f2} shows, as normalized histograms,
numerical results for single series of $L$ realizations of the
stochastic process after $t=10^3$ steps, from four different initial
conditions. In (a), all the languages have exactly the same initial
size $p_0$. In (b), the initial sizes are uniformly distributed
between $p=0$ and $p_{\max}$. In (c), the distribution of initial
sizes is also uniform, but spans the interval $(p^*,2p^*)$. Finally,
in (d) the initial distribution is more heterogeneous, with half the
languages having size $p^\dagger$ and the other half having size $10
p^\dagger$. The parameters $p_0$, $p_{\max}$, $p^*$, and
$p^\dagger$, which characterize these initial distributions, are
fixed by the condition that the total population is $P_0=3.1 \times
10^8$. The curve in all plots is the Gaussian of Eq. (\ref{Gauss})
with the parameters of Eq. (\ref{fit}). The agreement in cases (a)
to (c) is excellent. Only in case (d), where the initial
distribution is specially heterogeneous --and, certainly, not a
likely representation of the distribution of language sizes ten
centuries ago-- the deviations are larger, though the agreement is
still very reasonable.

The difference between the numerical results of case (d) and the
expected Gaussian function resides not only in the width of the
distribution but also in its mean value. To a much lesser extent,
this discrepancy is also visible in case (b). This shift between the
distribution peaks can be understood in terms of the average of Eq.
(\ref{qt}) over both the realizations of the growth rate
$\alpha_t^{(i)}$ and the initial log-sizes $q_0^{(i)}$,
\begin{equation} \label{mq}
\langle q_t^{(i)}\rangle = \langle q_0^{(i)}\rangle+ t( \ln \langle
\alpha \rangle-\sigma_\delta^2/2).
\end{equation}
Besides the contribution of the multiplicative stochastic process,
given by the term proportional to the time $t$, the mean value
$\langle q_t^{(i)}\rangle$ in Eq. (\ref{mq}) depends on the average
initial log-size $\langle q_0^{(i)}\rangle$. In spite of the fact
that the total initial population and the number of languages are
the same for all simulations --which always gives the same average
size per language-- the average log-size depends on the specific
initial distribution. Thus, the final mean values for different
initial conditions are generally shifted with respect to each other.

As a consistency test for the assumption that languages do not die
out along the evolution period considered here, I have also run
simulations taking into account the absorbing boundary at $p=1$.
Namely, when the size of a language decreases below one speaker
during its evolution, it is considered to become extinct and that
particular realization of the stochastic process is interrupted.
Among the four initial conditions considered above, those who
undergo larger extinctions are, not unexpectedly, (b) and (d) --as
they produce the largest shifts to the left in the log-size
distributions. In both cases, however, the fraction of extinct
languages is around 1~\%, which validates the above assumption quite
satisfactorily.

\subsection{Size distribution within language families}

A crucial consequence of the hypotheses on which the present model
is based --in particular, the mutual independence of the size
evolution of different languages-- is that its predictions should
hold not only for the ensemble of all the world's languages, but
also for any sub-ensemble to which the homogeneity assumptions (i)
and (ii)  reasonably apply. In other words, the final log-normal
shape of the size distribution should also result from the evolution
of, say, the languages of a given geographical region, or belonging
to a given language family. This can be readily assessed from
empirical data on the number of speakers of individual languages
and, in fact, has already been pointed out for a set of some 1,000
New Guinean languages.\cite{novotny}

\begin{figure}[h]
\centerline{\psfig{file=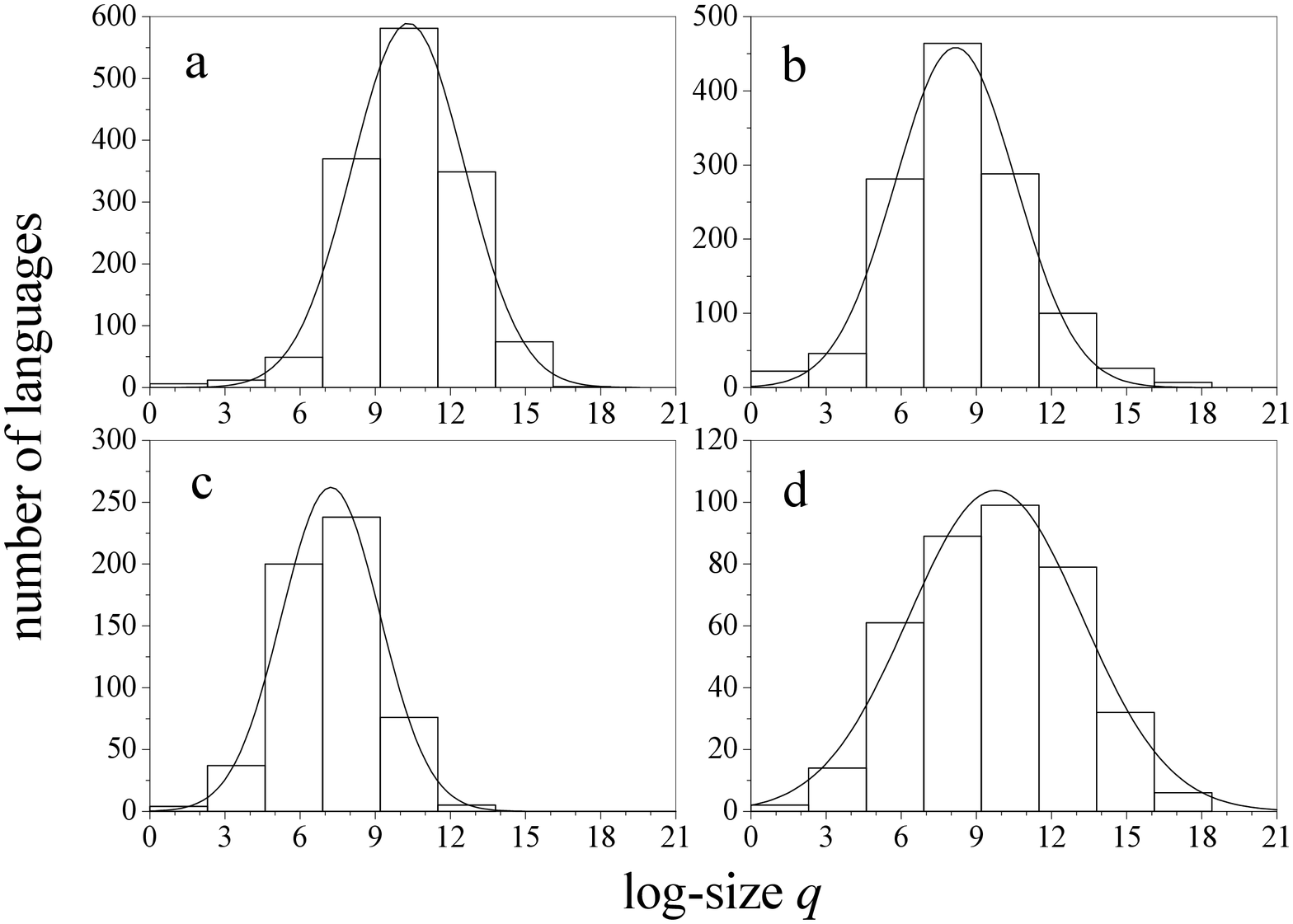,width=10cm}} \vspace*{8pt}
\caption{Histograms of the number of languages as a function of the
language log-size for four language families: (a) Niger-Congo (1,495
languages), (b) Austronesian (1,246 languages), (c) Trans-New Guinea
(561 languages), and (d) Indo-European (430 languages). Curves are
Gaussian least-square fittings.\label{f3}}
\end{figure}

Here, I analyze the size distribution of languages belonging to each
one of the four largest families, according to Ethnologue's
classification.\cite{ethnologue} Population data for individual
languages were obtained from Ethnologue's online databases. Figure
\ref{f3} shows histograms of log-sizes for those families. To ease
the comparison, the column width and the horizontal scale are the
same as in Fig. \ref{f1}. Curves stand for least-square fittings
with Gaussian functions as in Eq. (\ref{Gauss}). The resulting
parameters are $\bar q= 10.3$ and $\theta = 2.3$ for Niger-Congo;
$\bar q= 8.2$ and $\theta = 2.4$ for Austronesian; $\bar q= 7.2$ and
$\theta = 2.0$ for Trans-New Guinea; and $\bar q= 9.8$ and $\theta =
3.5$ for Indo-European. The quality of the agreement between the
data and the Gaussian fitting is clearly comparable to that of the
whole language ensemble in Fig. \ref{f1}.

Note the interesting fact that the mean square dispersion $\theta$
--which, according to the present model, results from the dispersion
in the population growth rate-- is sensibly larger for the
Indo-European family than for the other three. Surely, this is a
direct consequence of the highly diverse fate of European languages
in the last few centuries. In any case, the four mean square
dispersions are not far from the overall value given in Eq.
(\ref{fit}).

\section{Conclusion}

In this paper, I have argued that the present log-normal
distribution of language sizes is essentially a consequence of
demographic dynamics in the population of speakers of each language.
In fact, an isolated population can largely vary in number within
time scales which are short as compared with those involved in
substantial language evolution. To support this suggestion, I have
proposed a stochastic multiplicative process for the population
dynamics of individual languages, where language birth and death are
disregarded. Within some bold assumptions on the geographical  and
temporal homogeneity of the process, the model is completely
specified by two parameters, which give the average growth rate of
the population and its mean square dispersion. The average growth
rate is completely defined by the initial and the final world
population. I have chosen to apply the model on the period spanning
the last 1,000 years, for which reliable data on the world's
population growth are available. The mean square dispersion of the
growth rate is the only parameter which I fitted in an admittedly
{\it ad-hoc} manner, using the present distribution of language
sizes. It seems unlikely to find estimations for this parameter from
independent historical sources, which would require to have reliable
records on the population change year by year. Note that the
dispersion in the growth rate is determined by a variety of factors,
including fluctuations in birth and mortality frequencies and
migration events.\cite{sol}

Once the two parameters are fitted, the model is able to produce, as
the result of evolution along ten centuries, excellent predictions
of the present distribution of language sizes. Numerical simulations
show that the final distribution is largely independent on the
initial condition. This emphasizes the point that, irrespectively of
the long-range historical processes that may have determined the
distribution of language sizes of 1,000 years ago --including
language birth and death, branching, mutation, competition,
assimilation, and/or replacement-- population dynamics is by itself
able to explain the present distribution. This conclusion had
already been advanced for the case of New Guinean languages in Ref.
\refcite{novotny}. In fact, realizing that the same log-normal
profile is found in the size distribution inside language families,
is a further validation of the present model. In view of the present
arguments, one can moreover safely assert that the distribution of
language sizes was already a log-normal function, with different
parameters, in year 1000.

It is clear that in the last few years --with the advent of a host
of new mechanisms of globalization which endanger cultural
diversity-- many, or most, of the world's languages are threatened
by the risk of extinction.\cite{strog,sutherland,krauss} This risk
is peculiarly acute for those languages whose number of speakers is
below a few hundreds --including the range where the distribution of
sizes differs from the log-normal profile (cf. Fig. \ref{f1}). It
should be a program of obviously urgent interest to study in detail
what are the relevant processes at work in that range, even if they
escape the domain of the statistical physicists' approaches.

\section*{Acknowledgments}
I am grateful to Susanna C. Manrubia, who brought to my attention
the significance of the distribution of language sizes to the
problem of language evolution in 1996. Also, I acknowledge  critical
reading of the manuscript by A. A. Budini.

\end{document}